\documentclass[a4paper,11pt]{article}
\pdfoutput=1 

\usepackage{jinstpub} 
\usepackage{siunitx}
\usepackage{lineno}

\title{\boldmath Plasma density profile reconstruction of a gas cell for Ionization Induced Laser Wakefield Acceleration
}

\author[a]{F. Filippi,\note{Corresponding author.}}
\author[b]{L.T. Dickson,}
\author[c]{M. Backhouse, }
\author[d]{P. Forestier-Colleoni, }
\author[e]{C. Gustafsson, }
\author[f]{C. Cobo, }
\author[b]{C. Ballage, }
\author[d]{S. Dobosz Dufrénoy, }
\author[e]{E. Löfquist, }
\author[b]{G. Maynard, }
\author[f]{C.D. Murphy, }
\author[c]{Z. Najmudin, }
\author[a]{F. Panza, }
\author[e]{A. Persson, }
\author[a]{M. Scisci\'o, }
\author[b]{O. Vasilovici, }
\author[e]{O. Lundh, }
\author[b]{B. Cros}

\affiliation[a]{ENEA. Centro Ricerche Casaccia, 00123, Roma, Italia}
\affiliation[b]{LPGP, CNRS Université Paris-Saclay, Bat 210, 91405 Orsay, France }
\affiliation[c]{The John Adams Institute for Accelerator Science, Imperial College London, London, SW7 2AZ,UK}
\affiliation[d]{LIDYL, UMR9222, CEA, CNRS, Université Paris Saclay}
\affiliation[e]{Department of Physics, Lund University, P.O. Box 118, SE-22100, Lund, Sweden}
\affiliation[f]{York Plasma Institute, Department of Physics, University of York, Heslington YO10 5DD, UK}

\emailAdd{francesco.filippi@enea.it}

\abstract{Laser-driven plasma wakefields can provide hundreds of MeV electron beam in mm-range distances potentially shrinking the dimension of the actual particle accelerators. The plasma density plays a fundamental role in the control and stability of the acceleration process, which is a key development for the future electron injector proposed by EuPRAXIA.
A gas cell was designed by LPGP and LIDYL teams, with variable length and backing pressure, to confine the gas and tailor the gas density profile before the arrival of the laser. This cell was used during an experimental campaign with the multi TW-class laser at the Lund Laser Centre.  Ionization assisted injection in a tailored density profile is used to tune the electron beam properties. During the experiment, we filled the gas cell with hydrogen mixed with different concentration of nitrogen. We also varied the backing pressure of the gas and the geometrical length of the gas cell. We used a transverse probe to acquire shadowgraphic images of the plasma and to measure the plasma electron density. 
Methods and results of the analysis with comparisons between shadowgraphic and interferometric images will be discussed.
}


\begin{document}
\maketitle
\flushbottom

\section{Introduction}
Laser Wakefield Acceleration (LWFA) relies on the interaction of a short, high intensity laser pulse with a plasma which can be generated, for example, by gas discharge confined inside a capillary \cite{Spence2000, Filippi2018} or by laser ionization of neutral gas, properly shaped by a gas jet \cite{Semushin2001} or a gas cell \cite{Audet2018}. This technique can provide hundreds of \si{\mega\electronvolt} electron beam in mm-range distances, potentially shrinking the dimension of the actual particle accelerators, enhancing their impact on the scientific and industrial community. Currently, the produced electron beams are still far from the narrow energy spread and small emittance requested for the European project EuPRAXIA \cite{Eupraxia2020}. Many efforts are devoted to study mechanisms able to control the electron beam properties. A proposed method is the ionization assisted injection \cite{Lee2018}. This scheme uses the large difference in ionization potentials between successive ionization states of atoms to create electrons at a selected phase of the wakefield, resulting in low emittance beams. To achieve it, a mixture of gas with a hydrogen and a small percentage of nitrogen dopant has been used. A tailored density profile allows further tuning of the energy of the electrons and plays a fundamental role in the control and stability of the acceleration process \cite{Dickson2022}.

In this work, we will show the method we have used to acquire and analyse both shadowgrams and interferograms. 
This setup has been used for a LWFA experiment based on the ionization assisted injection technique to detect the plasma density produced in a gas-cell. Interferometry allows for quantitative measurements of the local plasma electron density and its spatial distribution. The phase shift caused by the plasma is expressed by the following formula \cite{Hutchinson2005}
\begin{equation}\label{eq:interferometry}
\Delta \phi = \frac{2\pi}{\lambda}\int_{L} 1-\left(1-\frac{n_e(z)}{n_c}\right)^{\frac{1}{2}} dz
\end{equation}
where $n_e$ is the free electron density, $L$ is the length of the path that the laser has crossed through the plasma, $\lambda$ is the laser wavelength and $n_c(\lambda)$ is the plasma critical density, proportional to $\lambda^{-2}$. We named $z$ the propagation axis along the probe. The phase shift can be made evident by introducing a small misalignment into the interferometer to have a pattern of linear interference fringes on the detection plane. The so-formed image can be recorded by a charge coupled device (CCD) which will allow the further analysis. Shadowgraphy, on the other hand, is sensitive to the spatial variation of the plasma density \cite{Hutchinson2005}
\begin{equation}\label{eq:shadowgraphy}
\frac{\Delta I}{I} = L \left[\frac{d^2}{dx^2} + \frac{d^2}{dy^2}\right] \int_L n_e(z) dz
\end{equation}
where $I$ and $\Delta I$ are the intensity and the intensity variation of the image on the CCD, respectively, and $x$ and $y$  are the coordinates transverse to the probe propagation axis. By combining the informations which can be given by the shadowgraphic and interferometric images it is possible to improve the plasma density reconstruction.
 
\section{Experimental setup}
The experiment has been performed at the Lund Laser Center. We used a \SI{20}{\tera \watt} Ti:Sa laser with a nominal duration of \SI{40}{\femto \second} FWHM, around \SI{1}{\joule} of pulse energy on target and \SI{800}{\nano \meter} central wavelength. The laser beam was focused with a $F/13$ 
off-axis parabola into the front surface of the gas cell. 
We used the gas cell ELISA (ELectron Injector for compact Staged high energy Accelerator), developed by LPGP and LIDYL teams \cite{ELISA}, which has been developed to confine and tailor the gas density profile before the arrival of the laser. The gas filling process was characterized both experimentally and by fluid simulations \cite{Audet2018}. 
The gas cell consists in a \SI{20}{\milli \meter} diameter cylinder with a \SI{3}{\milli \meter} diameter gas inlet located close to the cell entrance. Two replaceable steel plates with few hundreds microns thickness and \SI{400}{\micro \meter} diameter holes may be moved to modify the cell length from few hundred microns up to \SI{10}{\milli \meter}.
Optical quality windows, made of fused silica with-anti reflection coatings, are present on  each side for optical diagnostics transverse to the laser propagation axis. The two steel plates do not allow to probe the plasma at the entrance and the exit of the gas cell, then, only a small portion of the plasma can be probed.  
The gas inlet is connected to a reservoir and isolated by an electro-valve. The pressure inside the reservoir (backing pressure) dictates the neutral gas density inside the gas cell. During the experimental campaign, the backing pressure has been varied from \SI{0} up to \SI{300}{\milli \bar}.
We used the gas cell to confine hydrogen gas with a small percentage of nitrogen. During the experiment, we varied the backing pressure of the gas cell and we used different dopant concentrations, from \SI{0.25}{\percent} up to \SI{1}{\percent} of nitrogen dopant. 
Due to the large difference in the ionization potential between the inner K-shell of the nitrogen and the outer L-shell, hydrogen and the outer shell of the nitrogen can be ionized by the leading edge of the laser, contributing to the generation of the plasma wake and to the relativistic self-focusing, while the nitrogen K-shell electrons are ionized only nearby the laser peak intensity at a favorable wake phase, so the electrons can be immediately trapped by the wakefield. This mechanism is often named as ionization assisted injection.

\begin{figure}
	\centering
	a) \includegraphics[width=0.5\linewidth]{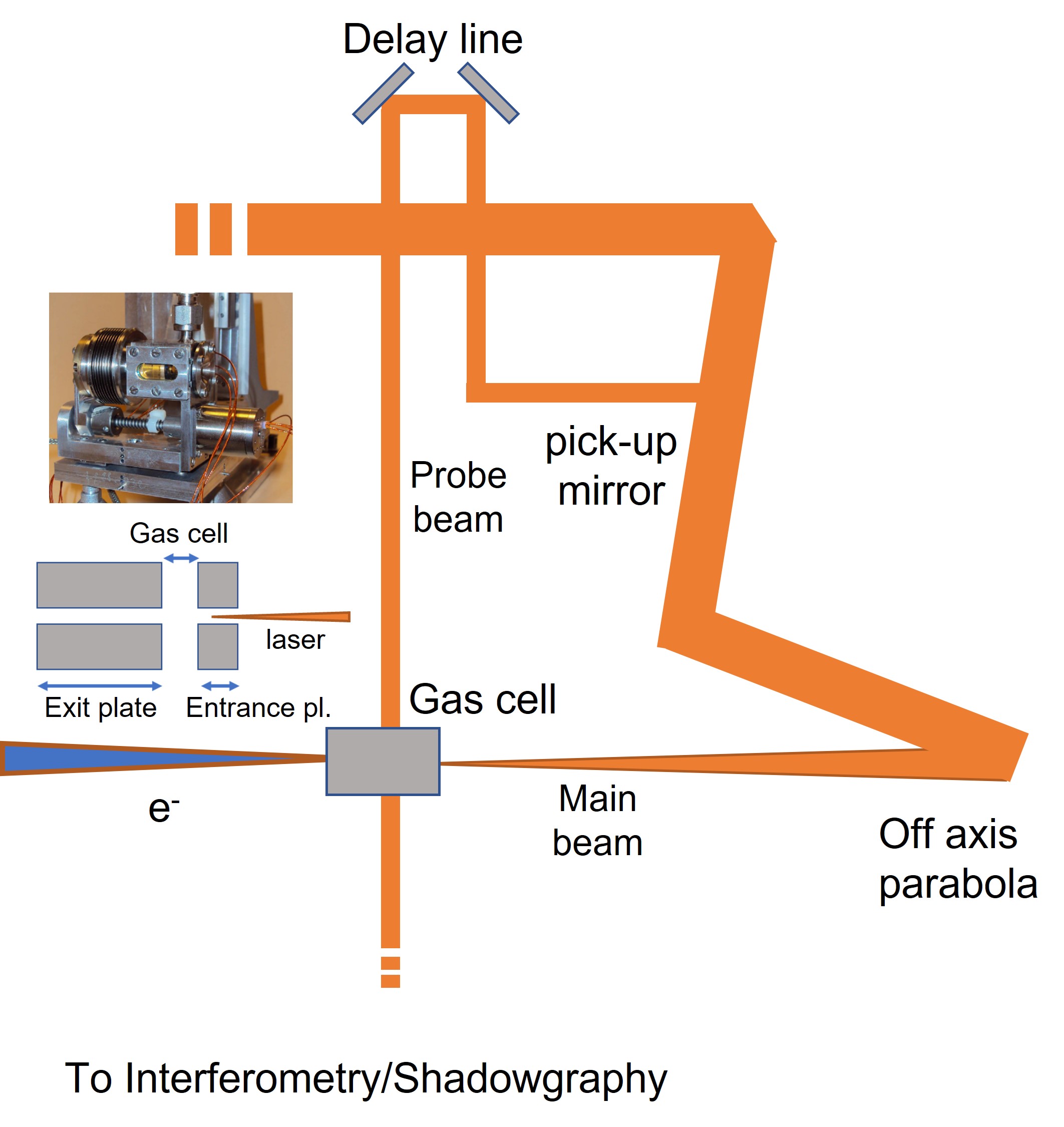} \hspace{1 cm} 
	b) \includegraphics[trim={0 2.5cm 0 0},width=0.22\linewidth]{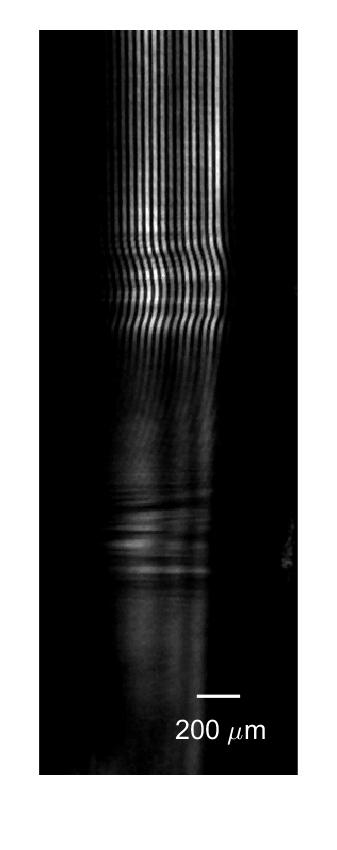} 
	\caption{On the left, there is a picture of the setup of the experiment with the probe beam path inside the experimental chamber. On the right, there is an example of the acquired data.}
	\label{fig:MZSetup}
\end{figure}

The background plasma density, i.e. the plasma contributing to the formation of the wakes, has beeen measured shot-by-shot with a probe beam which crossed the gas cell transversely immediately after the passage of the main beam. The geometry of the gas cell let to measure only a small portion of the plasma ionized by the laser. 
The probe has been obtained by moving a pick-up mirror into the main beam before the final focusing parabola deviating a small portion of the main laser. The timing between the main and the probe has been adjusted by moving a delay line inside the vacuum chamber. The interferometer used is a Mach-Zehnder type, placed outside the vacuum chamber. A schematic of the probe line inside the vacuum chamber and of the is shown in Fig. \ref{fig:MZSetup} a. The reference and the perturbed beams are obtained by splitting the same beam in two arms. The upper part of the image, with the plasma density, has been overlapped with the lower part, where the plasma is not present. 
With this setup, it is possible to image on the same CCD both the interferometric and the shadowgraphic images, which have crossed the same optical system, as shown in Fig. \ref{fig:MZSetup} b. With this schematic, there is only a single beam inside the vacuum chamber, avoiding the complications of a reference arm close to the gas cell. 

\section{Routine and analysis}
An accurate reconstruction of the plasma profile is of crucial importance both for the experimental analysis and for the simulations. 
We can divide our analysis in three steps. In the first step, we select the region of interest (ROI) for the interferometry of the acquired image, which is similar to the one in Fig. \ref{fig:MZSetup} b. The ROI must include the part of the image where the fringes bend, highlighting the presence of plasma. It should also include a region where there is no plasma and the fringes are straight. This will help the further part of the analysis since it gives a reference where to start to measure the shift. 
In the second step, we measure the phase shift of the fringes selected in the ROI. For doing this, we use the Fast Fourier Transform (FFT) method as described by \cite{Takeda1982}. In the third and final step, we retrieve the plasma density distribution from the measured phase shift. Since eq. \ref{eq:interferometry} belongs to the family of the Abel integrals, from the phase shift it is possible to obtain the plasma density distribution with a so-called Abel inversion if axially symmetric plasma can be assumed. Then, the output is a 2D circular section of plasma density, which can give the 3D distribution by rotating it along its central axis. 
Cylindrical symmetry is usually a good approximation for LWFA plasmas, nevertheless, the geometrical information required by the inversion algorithm should be given manually and Abel inversion is extremely sensitive to them. The input data are the position of the plasma rotational axis and the edges of the plasma column. Shadowgraphy is very sensitive to these plasma features \cite{Hutchinson2005, DeIzarra2012}. In the shadowgram, the edges of the plasma and its central axis are more evident, since they are highlighted by an intensity variation in the image. In the setup we have implemented, the plasma imaged in the shadowgram have the same dimensions of the one in the interferogram. Then a ROI with the same pixel dimensions of the one taken from the interferogram could be used in the shadowgraphy. By observing the shadowgraphy, it is possible to measure the distance between the axis and the plasma edges, and use them to analyse the interferogram. 
The main geometrical features of the plasma are taken from the shadowgraphy, where they are more evident, and applied to the interferographic image. This reduces the error in setting the rotation axis and the plasma edges, then improving the analysis.

\begin{figure}
	\centering
	a)\includegraphics[width=0.7\linewidth]{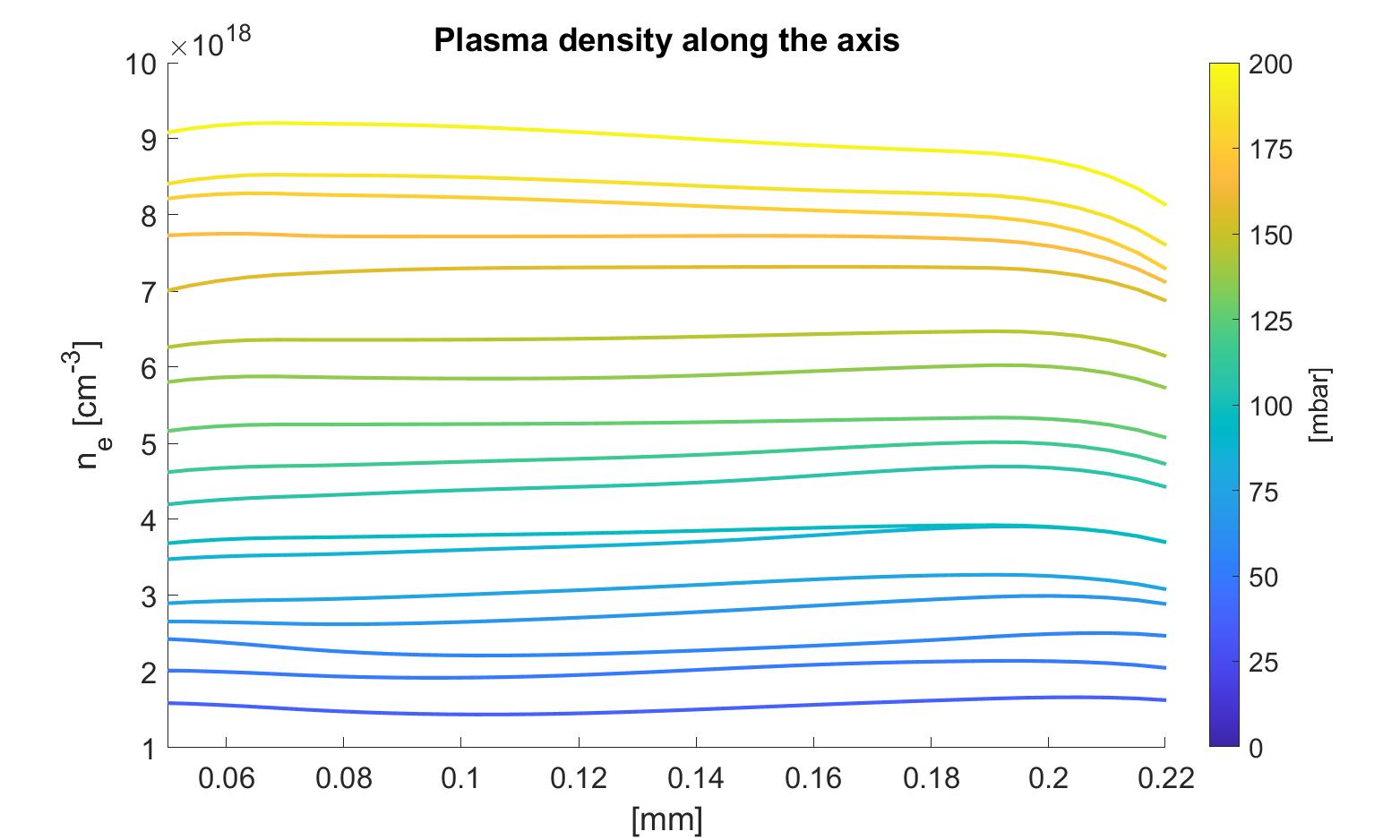} \\
	
	\bigskip
	
	b)\includegraphics[width=0.7\linewidth]{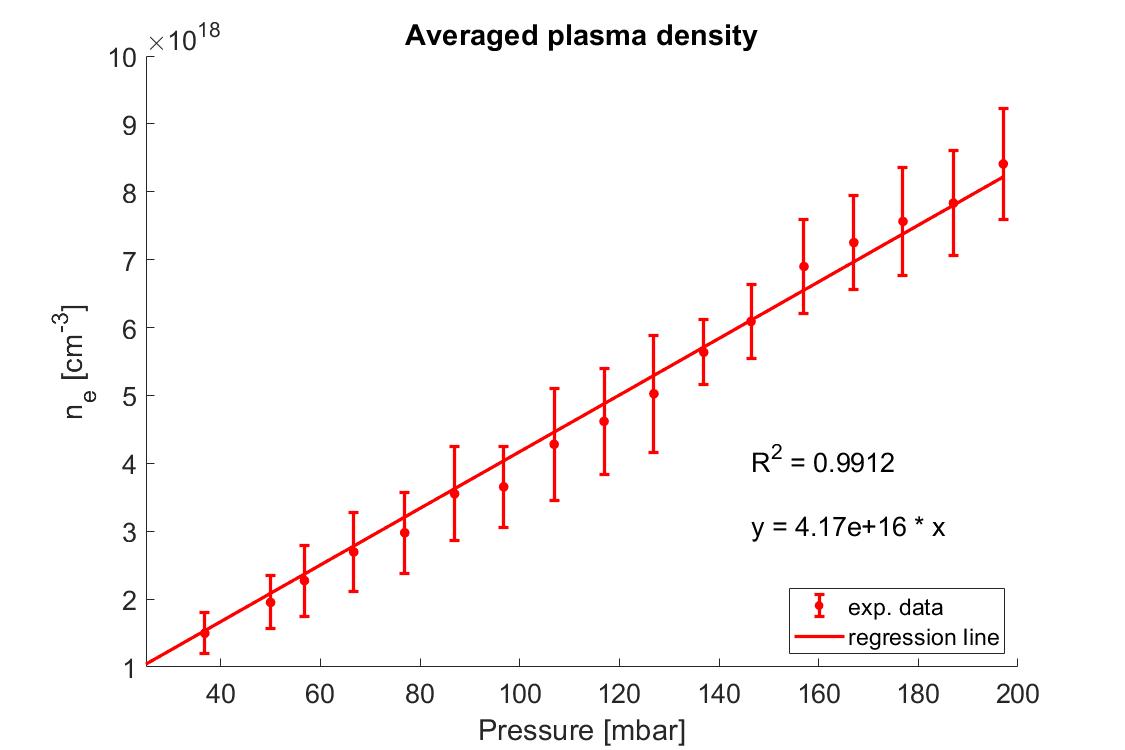} 
	\caption{In figure a), on the left, it is plotted the plasma density along the central axis of the plasma. We retrieved the plasma density distribution along the visible region of the gas cell, then we averaged the plasma density over \SI{50}{\micro\meter} radius around the plasma rotational axis. The longitudinal dimension is taken from the inner entrance face of the gas cell. The backing pressure has been varied from \SI{35} up to \SI{200}{\milli \bar}. In figure b), on the right, the mean value of the measured plasma density (with relative standard deviation) versus the set backing pressure is plotted. }
	\label{fig:lineprofile}
\end{figure}

We used this analysis to measure the plasma density in the gas-cell for different backing pressure of hydrogen doped with \SI{1}{\percent} of nitrogen. 
For those shots the gas cell had a \SI{500}{\micro \meter} thick entrance plate, less than \SI{300}{\micro \meter} long gas cell and \SI{2000}{\micro \meter} thick exit plate. Both the entrance and the exit orifices had a diameter of \SI{400}{\micro \meter}. Due to the geometry of the cell, only part of the gas cell can be imaged by the probe laser.
We varied the backing pressure from \SI{36}{\milli \bar} to \SI{200}{\milli \bar}. The results are shown in Fig. \ref{fig:lineprofile}. A homogeneous longitudinal plasma density were measured in the observed area, as can be seen in Fig. \ref{fig:lineprofile} a). In this figure, we retrieved the plasma density with the described method, then we averaged the plasma density over \SI{50}{\micro \meter} around the rotational axis. The plotted density is the longitudinal density profile of the plasma column along the laser propagation axis. As expected, the density increases with the increasing pressure. The same results have been averaged also longitudinally along the length of the plasma column, as shown in Fig. \ref{fig:lineprofile} b). The regression line fit well with the data and it can give a good reference to set the plasma density in further experimental campaigns. 

\section{Conclusions and perspectives}

We have described the setup of the shadowgraphic and interferometric system we have implemented for a LWFA experiment based on the ionized assisted injection technique. This setup allowed to acquire both the shadowgram and the interferogram on the same imaging plane, keeping information from both to optimize the plasma density reconstruction. We described the main steps of the analysis and the results for a pressure scan in the gas-cell used for the experiment have been shown. The results will be used for future experimental campaign to set the plasma density for the experiment, but will be also used to reproduce the experimental conditions for simulations.

Comparison between interferometric analysis and shadowgraphy will lead to more precise results. In the future, we want to further investigate the possible connections between those diagnostics, which may lead to a fast and operator-independent analysis for the plasma density reconstruction.

\section*{Acknowledgment}
This project has received funding from the European Union’s Horizon 2020 Research and Innovation programme under GA no 730871. 


\end{document}